\documentclass{Interspeech2024}

\usepackage{amsmath,graphicx, verbatim}
\usepackage{xcolor}
\usepackage{multirow}
\usepackage{float}
\usepackage{makecell}
\usepackage{amssymb}
\usepackage{algorithm}  
\usepackage{algpseudocode}  
\usepackage{pgfplots, hyperref}
\usepackage[raggedrightboxes]{ragged2e}
\usepackage{cancel}
\usepackage{booktabs,multirow}
\usepackage{adjustbox}
\usepackage[font=small]{caption}
\usepackage{subcaption}
\usepackage{enumitem}
\usepackage{pifont}

\usepackage[
backend=biber,
style=ieee,
citestyle=numeric-comp,
maxbibnames=3,
maxcitenames=3,
doi=false,isbn=false,url=false,eprint=false
]{biblatex}
\defbibheading{bibliography}[\refname]{}

\interspeechcameraready

\title{EFFUSE: Efficient Self-Supervised Feature Fusion for E2E ASR in Low Resource and Multilingual Scenarios}

\name[affiliation={1,2}]{Tejes}{Srivastava}
\name[affiliation={2}]{Jiatong}{Shi}
\name[affiliation={2}]{William}{Chen}
\name[affiliation={2}]{Shinji}{Watanabe}

\address{
  $^1$ University of Chicago 
  $^2$ Carnegie Mellon University}
\email{tejess@uchicago.edu, \{jiatongs, wc4, swatanab\}@csß.cmu.edu}

\keywords{Self-supervised learning, multilingual speech recognition, low-resource speech recognition}

\begin{document}

\maketitle

\begin{abstract}
    Self-Supervised Learning (SSL) models have demonstrated exceptional performance in various speech tasks, particularly in low-resource and multilingual domains. Recent works show that fusing diverse SSL models could achieve superior performance compared to using one SSL model. However, fusing models increases the overall parameter size, leading to higher computational costs. We propose EFFUSE, a novel approach that uses a single SSL model to mimic the features of multiple SSL models via prediction, resulting in a lightweight framework with competitive performance. Our experiments show that EFFUSE outperforms individual SSL models in multilingual speech recognition tasks. Our best performing model achieves an average SUPERB score increase of 63.5 (6.3\%) from the SSL baselines in Multilingual Speech Universal PERformance Benchmark (ML-SUPERB), while decreasing parameter size on average by 317M parameters (49\%) from the fusion models. 
\end{abstract}

\section{Introduction}

\label{sec:intro}

Recent works have incorporated self-supervised learning (SSL) models into end-to-end (E2E) automatic speech recognition (ASR) systems as frontend feature extractors to capture more relevant features of corpora~\cite{Chang2021AnEO, 9893562, Yang2021SUPERBSP, KrishnaD2021UsingLS, 9801640, shi2023multi, chen2023joint, chang2023colld}. Despite the encouraging progress in models equipped with an SSL frontend, there are still many limitations when applying SSL features. One such restraint is that most SSL models are trained with corpora in a single English talker scenario. Thus, this setting may not align well with other languages or recording conditions. Previous works have shown that these models might not capture enough information for downstream tasks in other languages or recording environments \cite{sanabria2022measuring, MengCLL22, ZuluagaGmez2022HowDP}. 
As a result, methods such as feature fusion \cite{berrebbi22_interspeech}, adapters \cite{10095130, 9746223}, teacher-student
distillation \cite{Peng2021ShrinkingBR, 9747490, Peng2023DPHuBERTJD, lee22p_interspeech, Yang2021KnowledgeDF}, and continual training  \cite{DBLP:conf/interspeech/HuangFZL22, vandereeckt_eusipco2022, Hsu2021RobustW2, 9746594}, have been proposed, which have shown performance improvements for different target domains.
Given the diversity of speech, with different varieties 
of language and recording conditions, it is difficult to capture all of the relevant information. Feature fusion is especially effective as each SSL can capture unique, distinct information \cite{berrebbi22_interspeech, fearless}. As a result, performing feature fusion captures the wide diversity of the speech features, by utilizing all of multiple SSL models' representations.
After the fusion, the output feature is expected to carry more information related to the downstream task.

Based upon the powerful benefits of feature fusion, we propose an efficient self-supervised feature fusion (\emph{EFFUSE}) methodology to utilize features from multiple SSL models, while limiting the increase in parameters and latency.
Specifically, EFFUSE consists of a two-stage training strategy: (1)~\textbf{Fusion stage}: train the model with a fusion of self-supervised features, similar to \cite{berrebbi22_interspeech, fearless}. (2) \textbf{Prediction stage}: continue to train the downstream model by using only one SSL model's features to predict other SSL model representations by a minimal number of linear layers.

Our proposed EFFUSE method reduces the number of parameters by using a \textit{single} SSL model to mimic the benefits of multiple SSL models, thereby scaling down inference costs compared to the original feature fusion network \cite{berrebbi22_interspeech, fearless}. 
We extensively validate our framework's effectiveness in both the low-resource and multilingual speech-processing tasks. EFFUSE demonstrates an average reduction of 4.5 absolute (20\% relative) character error rate (CER) in the low-resource domain over two benchmarks (Yoloxital Mixtec and Totonac) and 2.4 absolute (6.6\% relative) CER in the multilingual domain over the Multilingual Speech Universal PERformance Benchmark (ML-SUPERB). Notably, there is an average relative decrease of 16\% in the real-time factor (RTF) during inference, compared to the conventional fusion model.

Our contributions can be summarized as follows: 
(1) we extensively explore the SSL fusion model in low-resource and multilingual scenarios;
(2) we propose a novel framework that employs one SSL model to predict the features of other SSL models in the fusion network during inference; (3) we demonstrate that our approach can improve the performance of downstream multilingual tasks while maintaining reasonable inference costs.

\begin{figure*}[t]
\begin{subfigure}{.48\textwidth}
  \centering
  \scalebox{0.53}{
  \includegraphics{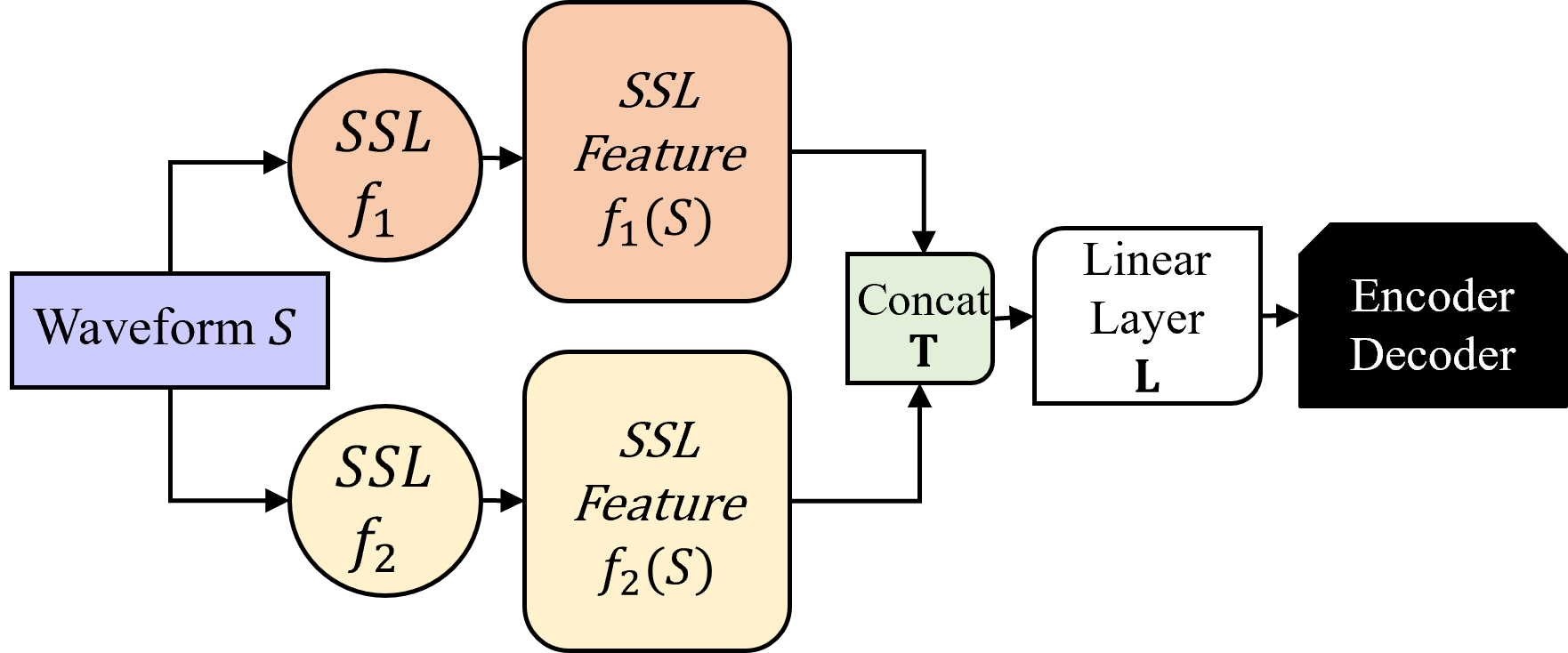}}
  \caption{Architecture of the model in the fusion stage. The raw waveform $S$ is passed through the SSL models $f_1, f_2$ to produce representations $f_1(S), f_2(S)$. These features are concatenated and projected as the input to the downstream ASR model.}
  \label{fig:sfig1}
\end{subfigure}%
\begin{subfigure}{.49\textwidth}
  \centering
    \scalebox{0.54}{
  \includegraphics{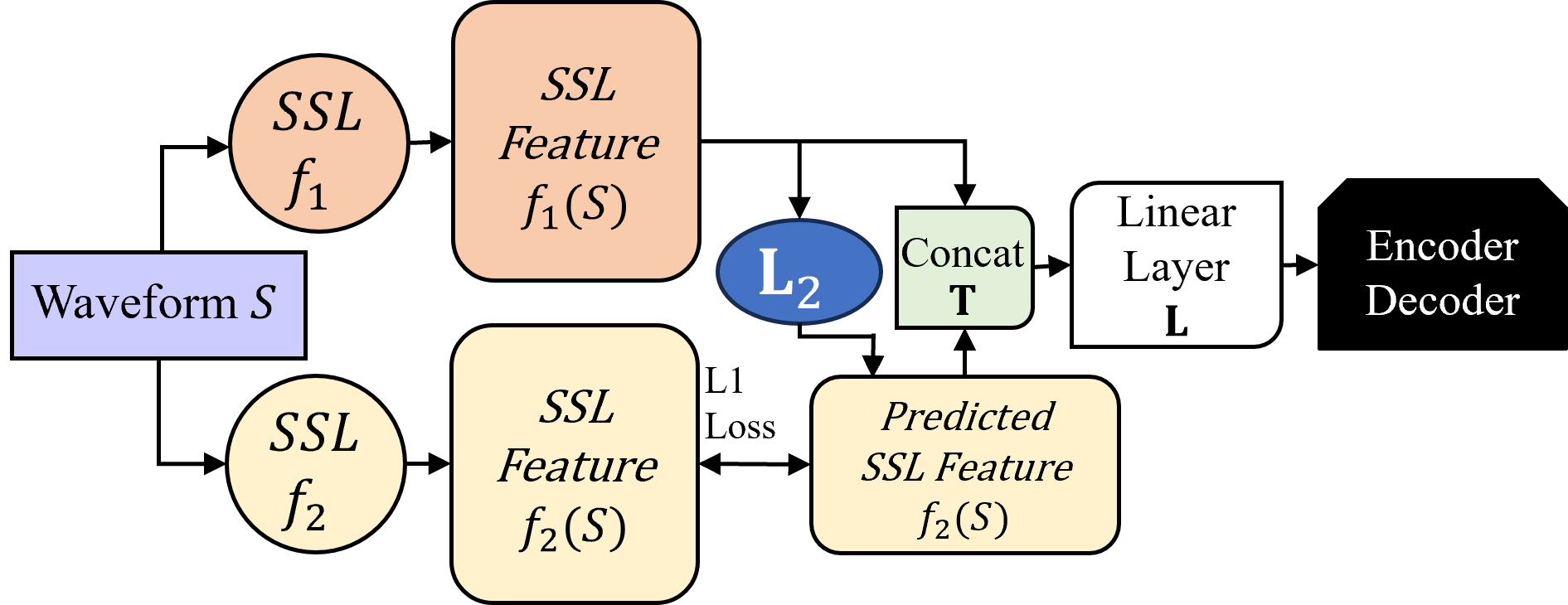}}
  \caption{Architecture of the proposed prediction stage. The raw waveform $S$ is passed through the SSL models $f_1, f_2$ to produce representations $f_1(S), f_2(S)$. The linear layer $\mathbf{L}_2$ is used to predict the representation $f_2(S)$. An L1-Loss function is used to train the linear layers. The features $ f_1(S)$, and $\mathbf{L}_2(f_1(S))$ are concatenated and projected to the ASR model. For inference, we can safely disregard SSL $f_2$.}
  \label{fig:sfig2}
\end{subfigure}
    \vspace{-5pt}
\caption{The proposed methodology with two SSL Models}
\vspace{-15pt}
\label{fig:fig}
\end{figure*}

\section{Predictive Ability Between SSL Models}
\label{ssec:predability}

\begin{table}[t]
\small
    \centering
    \caption{The coefficient of determination, $R^2$, is used to determine the correlation between model features. The $R^2$ values in the table represent how well the features from the left-side models predict weighted-sum features of the models on top.}
    \resizebox{0.9\linewidth}{!}{
    \begin{tabular}{c|c|c|c|c|c}
    \toprule
        & \multicolumn{5}{c}{Target} \\ \cmidrule{2-6}
        & & MFCC & HuBERT & WavLM & Wav2Vec~2.0\\ \midrule
        \multirow{5}{*}{\rotatebox[origin=c]{90}{Input}} & MFCC & $\times$ & 0.06 & 0.07 & 0.08\\ \cmidrule{2-6}
        & HuBERT & 0.67 & $\times$ & 0.71 & 0.60 \\ \cmidrule{2-6}
        & WavLM & 0.66 & 0.70 & $\times$ & 0.59\\ \cmidrule{2-6}
        & Wav2Vec2.0 & 0.68 & 0.63 & 0.65 & $\times$ \\ 
        \bottomrule
    \end{tabular}
    }
    \vspace{-15pt}
    \label{tab:r2table}
\end{table}

In this section, we discuss the capability of SSL models to predict other SSL models' features. This serves as a basis for the prediction 
model, which is the second stage of EFFUSE. 

In \cite{fearless}, the authors demonstrate a substantial linear correlation between features extracted from distinct SSL models in a fusion model. This observation paves the way for analyzing the predictive ability of SSL features. Given the complexities of considering different layers collectively, our attention is primarily drawn toward the weighted sum layer-wise features of the SSL models, which are widely used in SUPERB \cite{Yang2021SUPERBSP}.

To facilitate this, we train an ASR model on Totonac \cite{berrebbi22_interspeech}, a low-resource ASR corpus. The ASR model employs the weighted sum layer-wise features from frozen SSL models. Then, we extract weighted sum features of the pre-trained ASR 
model from 5,000 random utterances in the Totonac train set and use them to estimate linear regression models that predicts features of HuBERT, WavLM, and Wav2Vec~2.0, respectively. We select these three SSL models due to their diverse training objectives, which would, thus, assess the robustness of the predictability. 

To assess predictive capability, we employ the coefficient of determination (R$^2$) as a measure of variance explainability.\footnote{To calculate R$^2$, we utilize a matrix representation of the calculation since the target features have more than one dimension.} 
Our observations are presented in Table \ref{tab:r2table}, which exemplifies the SSL models' capabilities to predict weighted features.
The R$^2$ values are approximately 0.6 or higher, indicating a high correlation among features. Thus, we hypothesize that the prediction of features can retain the performance of fusion models to a great degree.

\section{EFFUSE Formulation}
\label{sec:method}

As discussed in Sec.~\ref{sec:intro}, EFFUSE contains two training stages, namely the fusion stage (Sec. \ref{ssec:fusion_model}) and the prediction stage (Sec. \ref{ssec: prediction_model}). In this section, we introduce the two stages of EFFUSE in details.

\subsection{Stage 1: Fusion}
\label{ssec:fusion_model}
We present this first stage in Fig. \ref{fig:sfig1}, and provide a detailed description of its inner workings as discussed in \cite{berrebbi22_interspeech, fearless}.
We begin by letting $S$ denote a sampled and quantized raw waveform. Let $\{f_1, \ldots, f_n\}$ be a set of $n$ SSL models. 
Suppose that for an SSL model $f_i$, there are $k_i$ layers. Then, for each layer, the SSL model extracts features from $S$ given by $\{f_i^{k}(S): k \in \{1, 2, \ldots, k_i\}\}$ where $f_i^k(S) \in \mathbb{R}^{r_i}$, where $r_i$ represents the dimension of the SSL model. 
The model learns weights for each layer to determine the importance of each layer to the downstream task. Using the learned weights for each SSL layer, the weighted sum features are computed as
\begin{equation}
\label{eqn:weightedsum}
    f_i'(S) = \sum_{k=1}^{k_i} w_i^{k}f_i^{k}(S).
\end{equation}
where $w_i^k$ is the weight of the $k$th layer.
We derive the set of weighted sum features as in Equation (\ref{eqn:weightedsum}) for $n$ SSL models as $\mathrm{F'} = \{f_1'(S), \ldots, f_n'(S)\}$. 

We note that each SSL model $f_i$ can have distinct resolutions and distinct output dimensions $r_i$. In order to perform feature fusion, we must unify these two parameters. A linear projection, followed by reshaping, on $\mathrm{F'}$ is performed so that all SSL features have a common number of frames and a common dimension $r$. Thus, we obtain features $\mathrm{F} = \{f_1(S), \ldots, f_n(S)\}$ where each $f_i(S) \in \mathbb{R}^r$. In Fig. \ref{fig:sfig1}, we denote two features: $f_1(S)$ and $f_2(S)$.

Now, let $\mathrm{\mathbf{T}}: (\mathbb{R}^{r})^n \to \mathbb{R}^{nr}$ denote a transform function, such as concatenation, linear, convolutional, and co-attention modules that fuse the different representations in $\mathrm{F}$, as described in \cite{berrebbi22_interspeech}. Then, let $\mathbf{L}\in \mathbb{R}^{nr\times m}$ symbolize a linear layer, where $m$ is the input dimension of the encoder-decoder model. In line with \cite{berrebbi22_interspeech}, the fusion feature $f_{\text{FUSE}}$ is represented as follows:
\begin{equation}
\label{eqn:sslfuse}
    f_{\text{FUSE}}(S)=\mathbf{L}(\mathrm{\mathbf{T}}(f_{1}(S), \ldots, f_n(S))) \in \mathbb{R}^{m},
\end{equation}
The fusion feature $f_{\text{FUSE}}$ is fed into an encoder-decoder architecture
as shown in Fig. \ref{fig:sfig1}. The model that uses this fusion procedure is termed the \textit{fusion model}.

\subsection{Stage 2: Prediction}
\label{ssec: prediction_model}

As discussed in Section \ref{ssec:predability}, SSL models possess the capability to predict the weighted sum features of other SSL models. In the prediction stage, we leverage this ability to develop a framework that preserves the exceptional performance achieved during the fusion stage. Importantly, this framework significantly reduces the parameter size by utilizing only one SSL model for inference.

As illustrated in Fig.~\ref{fig:sfig2}, the layer-wise weights $w_i^{k}$ (Eq. \eqref{eqn:weightedsum}) in each SSL model are frozen to focus on learning the weighted sum features. An SSL model in the set $\mathrm{F}$ is chosen as the prediction source, denoted as $f_1$. 
We then define $n-1$ linear layers given by  $\mathbf{L}_2, \ldots, \mathbf{L}_n$, where $\mathbf{L}_i \in \mathbb{R}^{r\times r}$.
Thus, in Fig. \ref{fig:sfig2}, there is one linear layer $\mathbf{L}_2$ as there are two SSL models.
Next, we utilize these linear layers to predict the weighted sum representations. Thus, a projected representation for the feature $f_k(S)$ is obtained as $\mathrm{\mathbf{L}}_k(f_1(S))$.
Then, the set of predicted features is

\begin{equation}
\label{eqn:sslrep}
    \{f_1(S), \mathrm{\mathbf{L}}_{2}(f_1(S)), \ldots, \mathrm{\mathbf{L}}_{n}(f_1(S)) \}.
\end{equation}
The predicted features in (\ref{eqn:sslrep}) are then fused, using a linear layer $\mathbf{L}$ and the same transform $\mathbf{T}$ in \eqref{eqn:sslfuse} and yield,
\begin{equation}
\label{eqn:sslpred}
\begin{aligned}
    f_{\mathrm{PRED}} = \mathbf{L}(\mathrm{\mathbf{T}}(f_1(S), \mathrm{\mathbf{L}}_{2}(f_1(S)), \ldots, \mathrm{\mathbf{L}}_{n}(f_1(S)))),
\end{aligned}
\end{equation}
where $f_{\mathrm{PRED}}$ denotes the final representation of the waveform $S$. 
Note that the difference between \eqref{eqn:sslfuse} and \eqref{eqn:sslpred} is that we are fusing weighted features of the SSL models directly in \eqref{eqn:sslfuse}. In \eqref{eqn:sslpred}, we specifically fuse only the predicted features.
Finally, to optimize the introduced linear layers ($\mathbf{L}_2, ..., \mathbf{L}_n$), we use the L1-loss function $\mathsf{Loss}_i = \mathrm{L1\_Loss}\{\mathrm{\mathbf{L}}_{i}(f_1(S)), f_i(S)\}$, 
where the loss is jointly trained with the downstream objective. The model using this second stage of EFFUSE is called the \textit{prediction model} or the \textit{EFFUSE model}.

\section{Experiments}
\label{sec:experiments}

\subsection{Datasets}
We investigate EFFUSE's effectiveness in the low-resource and multilingual scenarios. The low resource setting assesses EFFUSE's performance in an out-of-domain scenario with limited data. The multilingual domain tests the efficacy of EFFUSE when working with multiple languages.
In the low-resource setting, we experiment with two endangered languages with different typological characterisitcs: Totonac \cite{berrebbi22_interspeech} (a morphologically complex language) and Yoloxital Mixtec \cite{shi2021leveraging} (a tonal language).
For the multilingual setting, we utilize the dataset ML-SUPERB (public), a multilingual benchmark containing 240 languages with a 10 minute and 1 hour set \cite{shi2023mlsuperb, shi2023findings}.



\begin{table}[t]
\scriptsize
\caption{Results for Low Resource Domain. The fusion models yield optimal results across both datasets, while the prediction models demonstrate high-performance similar
to the fusion models with minimal parameter increase.}
    \vspace{-5pt}
\label{table:LRresults}
\centering
\resizebox {0.9\linewidth} {!} {
\begin{tabular}{l |c| c| c|c} 
\toprule
&  \multicolumn{2}{c|}{\textbf{YM}} & \multicolumn{2}{c}{\textbf{Totonac}} \\ 
Model & \multicolumn{1}{c}{WER} & \multicolumn{1}{c|}{CER}  & \multicolumn{1}{c}{WER} & \multicolumn{1}{c}{CER}  \\  \midrule
HuBERT (\texttt{H})  & 21.8 & 10.5 &   55.0 & 23.2  \\
WavLM (\texttt{WL}) & 21.0 & 10.4 & 52.3 & 21.7  \\
Wav2Vec~2.0 (\texttt{WV})& 20.5 & 9.9  & 56.4 & 26.3  \\ \midrule
\texttt{H+WL+WV} [Topline] & 20.3 & 9.6 & \textbf{45.2} & \textbf{14.5}  \\ \midrule
\texttt{H$\rightarrow$WV+WL} [Proposed] & 20.6 & 9.9  & 49.1 & 15.9  \\
\texttt{WL$\rightarrow$H+WV} [Proposed] & \textbf{20.2} & \textbf{9.6}  & 46.9 & 14.7 \\
\texttt{WV$\rightarrow$H+WL} [Proposed] & 20.3 & 9.8  & 46.2 & 15.2\\ \midrule \midrule 
MMS (\texttt{M})  & 22.6 & 10.4 & 48.2 & 16.6 \\
XLS-R (\texttt{X}) & 23.0 & 10.4 & 44.9 & 14.7 \\ \midrule
\texttt{M+X} [Topline] &  24.1 & 10.7 & \textbf{43.8} & 14.0 \\ \midrule
\texttt{M$\rightarrow$X} [Proposed]  & \textbf{21.8} & \textbf{9.8} & 46.2 & 15.4 \\
\texttt{X$\rightarrow$M} [Proposed] & 22.6 & 10.1 & 44.6 & \textbf{13.9} \\
\bottomrule
\end{tabular}
}
    \vspace{-15pt}
\end{table}

\subsection{Experimental Setup}
\label{ssec:setup}

All the experiments are conducted with ESPnet \cite{watanabe2018espnet} with S3PRL~\cite{Yang2021SUPERBSP} and its corresponding recipes with the four datasets.\footnote{All of our configurations, sources, and model checkpoints will be publicly released for reproducibility after the anonymization period.}

\noindent \textbf{SSL Models}: We select SSL models based on factors such as training objectives, and pretraining data domains. Our experiments focus on five widely used SSL models: HuBERT \cite{HuBERT}, WavLM \cite{WavLM}, Wav2Vec~2.0 \cite{Wav2vec2}, MMS \cite{pratap2023mms}, and XLS-R \cite{XLS-R}. We utilize the 300M parameter versions of these models.

\noindent \textbf{Baseline}: In our low-resource study, we choose HuBERT (\texttt{H}), WavLM (\texttt{WL}), Wav2Vec~2.0 (\texttt{WV}), MMS (\texttt{M}), and XLS-R (\texttt{X}) as the baseline models. We concentrate on conducting experiments with English-based SSL models to investigate EFFUSE in scenarios where there is minimal prior knowledge about the target domain, as observed in low-resource language settings. For ML-SUPERB, our baseline SSL models are MMS (\texttt{M}) and XLS-R (\texttt{X}). 

Both the YM and Totonac corpora use the same transformer encoder-decoder architecture as corresponding ESPnet recipes.  Speed perturbation \cite{Ko2015AudioAF} and SpecAugment \cite{SpecAugment} are used in low-resource scenarios.
For the ML-SUPBERB experiments, we follow the same setting as the framework design in \cite{shi2023mlsuperb}.



\noindent \textbf{Fusion Model}: 
In the low-resource setting, we utilize the combination of HuBERT, WavLM, and Wav2Vec~2.0 (\texttt{H+WL+WV}) together and also perform fusion with MMS and XLS-R (\texttt{M+X}) experiment. For the multilingual setting, we only fuse MMS and XLS-R (\texttt{M+X}). All of the features are fused as in Eq.\eqref{eqn:sslfuse}.

As discussed in  Sec. \ref{ssec:fusion_model}, we must select a transformation $\mathbf{T}$ and the dimension $m$ for $\mathbf{L}$. We choose $\mathbf{T}$ as concatenation and apply it over the features. Finally, we project the concatenated features using $\mathbf{L}$ into a 100-dimensional space. This procedure is uniformly applied to each feature across all the experiments we conduct.

\noindent \textbf{Prediction Model}: As discussed in Sec. \ref{ssec: prediction_model}, we select a prediction source ($f_1$ in Eq. \eqref{eqn:sslpred}) amongst the SSL models from the fusion model. 

For the fusion model in the low resource setting, there are three corresponding prediction models for English-based SSLs: (1) HuBERT as the prediction source [\texttt{H$\rightarrow$WV+WL}]
(2) WavLM as the prediction source [\texttt{WL$\rightarrow$H+WV}], and (3) Wav2Vec~2.0 as the prediction source [\texttt{WV$\rightarrow$H+WL}]
For the multilingual models, we have two prediction models: (4) MMS as the prediction source [\texttt{M$\rightarrow$X}] and (5) XLS-R as the prediction source [\texttt{X$\rightarrow$M}].

For ML SUPERB, both models that are utilized in the fusion model are used as prediction sources in two different experiments. Specifically, we have MMS as the prediction source (\texttt{M$\rightarrow$X}) and XLS-R as the prediction source (\texttt{X$\rightarrow$M}).

Throughout these experiments, we retain the architecture of the fusion models.

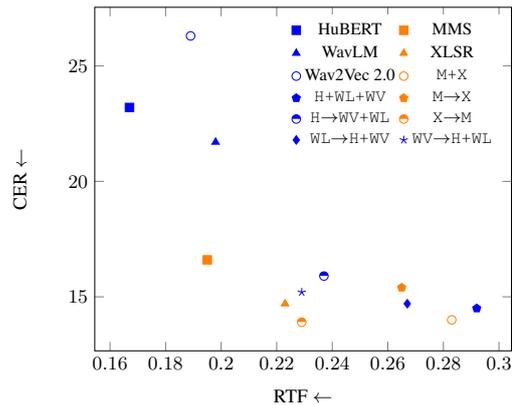
\begin{figure}[!t]
\centering
\begin{tikzpicture}[scale=0.8]
\begin{axis}[
    xlabel={RTF $\leftarrow$},
    ylabel={CER $\leftarrow$},
    legend style={at={(0.98,0.98)}, font=\footnotesize, anchor=north east, draw=none, legend columns=2},
    scatter/classes={
        a={mark=square*,blue},
        b={mark=square*,orange},
        c={mark=triangle*,blue},
        d={mark=triangle*,orange},
        e={mark=o,blue},
        f={mark=o,orange},
        g={mark=pentagon*,blue},
        h={mark=pentagon*,orange},
        i={mark=halfcircle*,blue},
        j={mark=halfcircle*,orange},
        k={mark=diamond*,blue},
        l={mark=star,blue}
    }
]
\addplot[scatter,only marks,
    scatter src=explicit symbolic]
    coordinates {
        (0.167,23.2) [a]
        (0.195,16.6) [b]
        (0.198,21.7) [c]
        (0.223,14.7) [d]
        (0.189,26.3) [e]
        (0.283,14.0) [f]
        (0.292,14.5) [g]
        (0.265,15.4) [h]
        (0.237,15.9) [i]
        (0.229,13.9) [j]
        (0.267,14.7) [k]
        (0.229,15.2) [l]
    };
\legend{HuBERT, MMS, WavLM, XLSR, Wav2Vec~2.0, \texttt{M+X}, \texttt{H+WL+WV}, \texttt{M$\rightarrow$X}, \texttt{H$\rightarrow$WV+WL}, \texttt{X$\rightarrow$M}, \texttt{WL$\rightarrow$H+WV}, \texttt{WV$\rightarrow$H+WL}}
\end{axis}
\end{tikzpicture}
\caption{RTF vs. CER for Totonac. The individual SSL models demonstrate excellent performance in terms of RTF, but there is a trade-off with CER. The prediction models strike a balance by maintaining high performance in both CER and RTF, with additional models included for comparison.}
\label{fig:CER_RTF}
\vspace{-15pt}
\end{figure}

\begin{table*}[!t]
\small
    \centering
        \caption{\{10-minute / 1-hour\} set ML-SUPERB benchmark. Both stages of EFFUSE yield reasonable improvements over the baselines and similar performance with the topline fusion model. Refer to more discussions in Sec.~\ref{ssec:multi}.}
            \vspace{-5pt}
    \resizebox {\linewidth} {!} {
\begin{tabular}{l|c|c|cc|c|ccc|c}
\toprule
\multirow{3}{*}{SSL} & \multirow{3}{*}{Info} & Monolingual ASR & \multicolumn{2}{c|}{Multilingual ASR} & \multicolumn{1}{c|}{LID} & \multicolumn{3}{c|}{Multilingual ASR + LID} & \multirow{2}{*}{SUPERB$_{s}$} \\
&          &       &      Normal & Few-shot & Normal & \multicolumn{2}{c}{Normal} & \multicolumn{1}{c|}{Few-shot} \\
& & CER $\downarrow$ & CER $\downarrow$ & CER $\downarrow$ & ACC $\uparrow$ & ACC $\uparrow$ & CER $\downarrow$ & \multicolumn{1}{c|}{CER $\downarrow$} & $\uparrow$  \\
\midrule
MMS (\texttt{M})& Baseline & \textbf{33.8} / 30.5 & 29.9 / 24.6 & 35.0 / 35.4 & 62.3 / 84.3 & 73.8 / 88.9 & 29.5 / 24.5 & 34.9 / 35.4 & 1051.7 / 1029.2 \\ 
XLS-R (\texttt{X})& Baseline & 39.5 / 30.5 & 28.9 / 21.6 & 41.4 / 39.1 & 65.4 / 87.2 & 76.9 / 90.4 & 28.7 / 22.0 & 41.3 / 38.3 & 957.6 / 1018.1 \\ \midrule \midrule
MMS+XLS-R (\texttt{M+X}) & Fusion [Topline] & 35.6 / \textbf{27.9} & 27.0 / \textbf{20.6} & 38.9 / 36.6 & 81.3 / 86.3 & 78.1 / 91.9 & 26.8 / \textbf{20.5} & 38.7 / 36.9 & 1095.2 / \textbf{1068.4} \\ \midrule \midrule
MMS$\rightarrow$ XLS-R (\texttt{M$\rightarrow$X}) & Prediction [Proposed] & 38.9 / 29.4 & \textbf{26.5} / 22.3 & \textbf{33.8} / \textbf{35.4} & 83.4 / 84.9 & \textbf{82.9} / \textbf{92.1} & \textbf{26.3} / 21.6 & \textbf{33.8} / \textbf{34.0} & \textbf{1152.3} / 1066.5 \\
XLS-R$\rightarrow$MMS (\texttt{X$\rightarrow$M}) & Prediction [Proposed]& 38.6 / 28.9 & 28.3 / 21.3 & 41.0 / 39.7 & \textbf{83.9} / \textbf{90.4} & 76.8 / 91.0 & 28.1 / 21.0 & 39.3 / 39.1 & 1054.8 / 1037.0 \\
\bottomrule
\end{tabular}
}
    \vspace{-15pt}
    \label{tab:combined_ml}   
\end{table*}

\begin{table*}[!t]
    \centering
        \caption{A summary of parameter efficiency. The table presents the increase in parameter size (rounded whole values) and SUPERB score with respect to the baselines with a single SSL frontend. The fusion and prediction models show an increase over their respective baselines. However, the parameter size increase
        for the prediction models is minimal compared to the fusion models.}
            \vspace{-5pt}
        \resizebox {0.8\linewidth} {!} {
    \begin{tabular}{c|c|c|c|c|c}
    \toprule
         Model & Num. Param. (M)  & \makecell{$\Delta$ Num. Param. (M) \\ (MMS Baseline)} &  \makecell{$\Delta$ Num. Param. (M) \\ (XLS-R Baseline)} & \makecell{$\Delta$ SUPERB$_s$ \\ (MMS Baseline)} & \makecell{$\Delta$ SUPERB$_s$ \\ (XLS-R Baseline)} \\ \midrule
         MMS (\texttt{M}) & 322 & / & / & / & /\\
         XLS-R (\texttt{X})& 324 & / & / & / & /\\ 
         \texttt{M+X} (10 min) & 641 & +319 & +317 & +43.5 & +137.6\\
         \texttt{M$\rightarrow$X} (10 min) & 324 & +2 & / & +100.6 & /\\
         \texttt{X$\rightarrow$M} (10 min) & 324 & / & +0 & / & +97.2\\
         \bottomrule
    \end{tabular}
 }
     \vspace{-15pt}
    \label{tab:param}
\end{table*}

\section{Results}
\subsection{Low Resource ASR}
\label{ssec:lowres}

Our results for our low resource study are summarized in Table~\ref{table:LRresults}. We highlight that the prediction models show improvements over the baselines.

\textbf{YM}: The \texttt{WL$\rightarrow$H+WV} demonstrates a decrease in WER while maintaining the CER from the English topline fusion model. Despite the drop in performance by the \texttt{M+X} model, we find that the multilingual EFFUSE models \emph{still outperform} the baselines, exemplifying EFFUSE's effectiveness.
\textbf{Totonac}:~\texttt{WL$\rightarrow$H+WV} exhibits the best performance for the English-based EFFUSE models, with minimal increase in CER from the corresponding fusion model. The \texttt{X$\rightarrow$M} shows similar performance to the multilingual fusion model.
Our results show that EFFUSE scales well as the number of SSL models increases.

Thereafter, we analyze the efficiency of inference in Figure~\ref{fig:CER_RTF} over the Totonac benchmark. We observe that the baseline models
achieve the fastest inference speeds but have substandard performance. The fusion model, on the other hand, significantly suffers
in terms of inference speed but yields the optimal CER. The prediction model functions as the most economical framework, retaining the 
performance of the fusion model. Importantly, these models experience a noteworthy speedup that is only slightly slower than the 
singular SSL-based ASR models.

\subsection{ML-SUPERB Results} \label{ssec:multi}


As introduced in \cite{shi2023mlsuperb}, there are four tasks (monolingual ASR, multilingual ASR, LID, multilingual ASR+LID) with two data configurations (i.e., 10-minute and 1-hour). The results for both sets with all tasks are recorded in Table~\ref{tab:combined_ml}.

\noindent
\textbf{Overall}: In the 10 minute set, we highlight that our \texttt{M$\rightarrow$X} model yields the best performance with a SUPERB score of \textit{1152.3}, even outperforming the topline \texttt{M+X} model. This result demonstrates that EFFUSE has the capabilities of surpassing fusion models. In the 1 hour set, we showcase that the \texttt{M+X} model yields a performance of 1068.4. We highlight that the \texttt{M$\rightarrow$X} model provides competitive performance with SUPERB score \textit{1066.5} with much less inference costs.

\noindent
\textbf{Monolingual ASR Track}: 
In the 10-minute set, we find that the baseline models give best performance. The fusion and EFFUSE models have mixed results, which might be attributed to the limited data size.
In the 1 hr dataset, our prediction models exceed the performance of the baseline models.

\noindent
\textbf{Multilingual ASR Track}: 
The EFFUSE models generally surpass the performance of their corresponding baselines. These models demonstrate the ability to outperform the fusion (topline) model as well, exemplified by the \texttt{M$\rightarrow$X} model. Even in the case when EFFUSE models do not exceed the fusion model's performance, they still usually show a significant improvement over the baselines.

\noindent
\textbf{LID Track}: 
The prediction models show improvements over their respective baselines, with their performance closer to the topline model. In fact, the EFFUSE models also exhibit that they are capable of outperforming the fusion model, as in the \texttt{X$\rightarrow$M} case. 

\noindent
\textbf{Joint Multilingual+LID Track}: The EFFUSE prediction models generally outperform the baselines.
The prediction models either outperform or yield similar performance to the fusion model. This indicates that the prediction models learn the representations of the fusion model fairly accurately.

\subsection{Computational Cost} \label{ssec: cost}

As shown in Sec. \ref{ssec:multi}, both stages of EFFUSE exhibit boosts over 
their corresponding baselines. Table \ref{tab:param} provides a summary of parameter sizes for the fusion and prediction models in ML-SUPERB (see Sec. \ref{ssec:multi}) and their impact on the SUPERB score.

We observe that the model \texttt{M$\rightarrow$X}  shows an increase of +100.6 in the SUPERB Score. This result is especially significant as with only a \textbf{+2M} parameter increase, it outperforms the corresponding fusion model, which has a parameter increase of +318M. Similarly, \texttt{X$\rightarrow$M} shows a +97.2 SUPERB score increase with +0M\footnote{This value is rounded down, as the parameter increase is minimal.}  parameter increase. With minimal parameter growth, these prediction models deliver competitive results that significantly outperform Baselines \texttt{X} and \texttt{M}.

\section{Conclusion}
We propose a simple, efficient methodology, EFFUSE, to leverage the predictive capabilities of SSL models. EFFUSE uses two stages: 1) the SSL fusion model is used as an initial training stage; 2) a prediction source is employed to successfully predict SSL-weighted features. The two stage can prove to be a hindrance, as training the fusion model is computationally expensive. However, upon evaluating our method over a multilingual benchmark and two low-resource domain benchmarks, we find the EFFUSE methodology can provide significant benefits in performance, while limiting parameter increase with a faster inference speed.

\section{Acknowledgements}
Experiments of this work used the Bridges2 system at PSC and Delta system at NCSA through allocations CIS210014 and IRI120008P from the ACCESS program, supported by NSF grants \#2138259, \#2138286, \#2138307, \#2137603, and \#2138296.

\section{References}
\printbibliography

\end{document}